\documentclass{article}

\usepackage{microtype}
\usepackage{graphicx}
\usepackage{subfigure}
\usepackage{booktabs} 

\usepackage{hyperref}

\usepackage[accepted]{icml2024}

\usepackage{amsmath}
\usepackage{amssymb}
\usepackage{mathtools}
\usepackage{amsthm}
\usepackage[capitalize,noabbrev]{cleveref}

\theoremstyle{plain}

\theoremstyle{definition}

\theoremstyle{remark}

\usepackage[textsize=tiny]{todonotes}

\icmltitlerunning{Quantum Embedding with Transformer for High-dimensional Data}

\begin{document}    

\twocolumn[
\icmltitle{Quantum Embedding with Transformer for High-dimensional Data}



\icmlsetsymbol{equal}{*}

\begin{icmlauthorlist}
\icmlauthor{Hao-Yuan Chen}{Hao-Yuan Chen}
\icmlauthor{Yen-Jui Chang}{Yen-Jui Chang}
\icmlauthor{Shih-Wei Liao}{Shih-Wei Liao}
\icmlauthor{Ching-Ray Chang}{Ching-Ray Chang}
\end{icmlauthorlist}

\icmlaffiliation{Hao-Yuan Chen}{Department of Computer Science, University of London, London, United Kingdom}
\icmlaffiliation{Yen-Jui Chang}{Department of Physics, National Taiwan University, Taipei, Taiwan}
\icmlaffiliation{Shih-Wei Liao}{Department of Computer Science and Information Engineering, National Taiwan University, Taipei, Taiwan}
\icmlaffiliation{Ching-Ray Chang}{Department of Physics, National Taiwan University, Taipei, Taiwan, Quantum Information Center, Chung Yuan Christian University, Taoyuan City, Taiwan}
\icmlcorrespondingauthor{Hao-Yuan Chen}{hc118@student.london.ac.uk}

\icmlkeywords{Quantum embedding, vision transformer, self-attention, kernel method}

\vskip 0.3in
]




\begin{abstract}
Quantum embedding with transformers is a novel and promising architecture for quantum machine learning to deliver exceptional capability on near-term devices or simulators. The research incorporated a vision transformer (ViT) to advance quantum significantly embedding ability and results for a single qubit classifier with around 3 percent in the median F1 score on the BirdCLEF-2021, a challenging high-dimensional dataset. The study showcases and analyzes empirical evidence that our transformer-based architecture is a highly versatile and practical approach to modern quantum machine learning problems. 
\end{abstract}

\section{Introduction}
\label{submission}
Quantum machine learning holds a promising potential for advancing the state-of-the-art machine learning model \cite{biamonte2017quantum}. Moreover, the research, \cite{lloyd2020quantum,gianani2022experimental}, has illustrated a theoretical and experimental framework for quantum embedding to deliver exceptional potential for advancing the state-of-the-art benchmark for various machine learning challenges. The research, \cite{lloyd2020quantum}, has introduced the initial concept of incorporating a classical convolutional neural network (CNN) to extract features from the various visual inputs. However, considering recent advancements in machine learning models like transformer \cite{vaswani2017attention} and vision transformer (ViT) \cite{mao2022towards} have made transformer models like ViT an ideal candidate for quantum machine learning on vast application domains.

Quantum embedding \cite{sun2016quantum}, a conceptual idea later validated as a practical technique in the research \cite{lloyd2020quantum} that a proper quantum kernel could radically improve the various machine learning solutions. Quantum machine learning \cite{schuld2021supervised} utilizes quantum embedding with the quantum kernel to project the data into Hilbert space. The expectation is to keep two separate classes far from two different clusters in the feature map, making effective and efficient classification possible. 

The research introduces a theoretical model for incorporating pre-trained transformer models trained for the ImageNet benchmark \cite{beyer2022better} to transform various visual inputs into linear representations of tensors for quantum embedding. The theoretical model introduced the concept of modeling various patterns using the self-attention \cite{46989} mechanism. Moreover, a proper mapping for the feature tensors to the quantum feature map is introduced to formulate the idea of quantum embedding. Ultimately, the empirical evidence from high-dimensional datasets like BirdCLEF 2021 for binary classification evaluates the method's effectiveness.

The research question concerns whether vision transformers are efficient and effective feature extraction and representation models for quantum embedding with the hybrid trainable approach. The empirical evidence of effectiveness is based on the binary classification metrics for quantum classifiers \cite{blank2020quantum}, including accuracy, precision, recall, and F1 score \cite{hossin2015review}. The result demonstrated a significant advancement in providing a universal embedding architecture for quantum neural networks on machine learning problems, particularly classification problems.

\section{Methods}   
\subsection{Model Architecture}
\begin{figure}[h]
\centering
\includegraphics[width=0.5\textwidth]{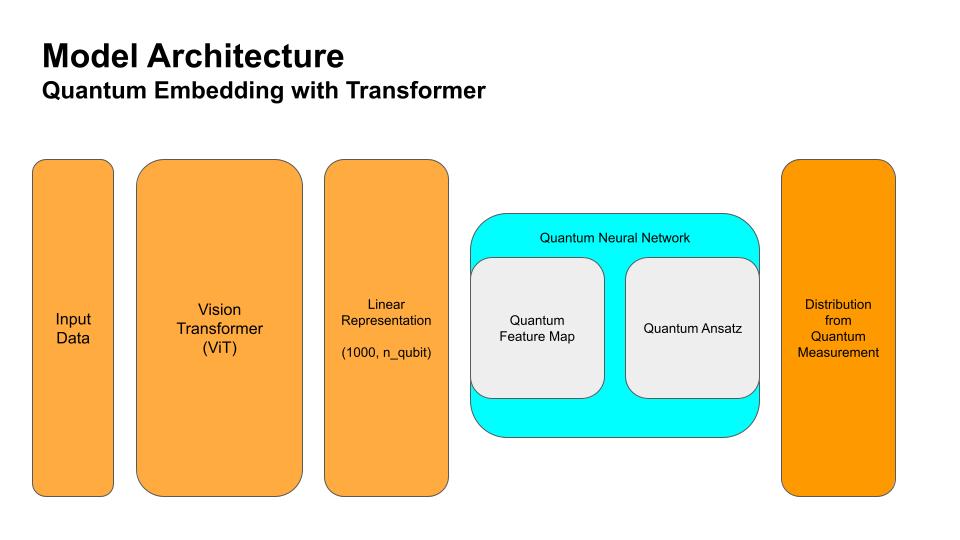}
\caption{Model architecture of the research on quantum embedding with transformers for high-dimensional visual datasets. The orange part stands for the classical state of data, and the blue part means a quantum state of information. The input data feeds from the left to the right to classify the input information.}
\label{fig:modelarchitecture}
\end{figure}

The embedding method in the research is illustrated in Figure \ref{fig:modelarchitecture}, which incorporates a transformer model with a linear representation layer to transform the original data input, i.e., images, to linear representation in tensors. Later, the linear features will be embedded into the quantum feature space with a quantum feature map using the trainable embedding technique \cite{Hubregtsen_2022}. By introducing a transformer model, like a vision transformer (ViT), the architecture provides a versatile, classical embedding layer for quantum feature maps to capture various data forms, including recurrent or sequential patterns.

\subsection{Vision Transformer}
The study facilitates Vision Transformer (ViT) to process and embed images using the Transformer architecture as follows:

\begin{enumerate}
    \item \textbf{Image Tokenization:} An input image \( I \) of size \( H \times W \times C \) is partitioned into a sequence of non-overlapping patches \( P \). Each patch of size \( N \times N \times C \) is flattened and linearly projected into a \( D \)-dimensional embedding space to create a sequence of patch embeddings \( X_p \).
    
    \item \textbf{Positional Encoding:} Positional encodings \( E_{pos} \) are added to the patch embeddings to provide positional information, resulting in the input sequence \( X = X_p + E_{pos} \).
    
    \item \textbf{Transformer Encoder:} The encoder consists of \( L \) layers, each containing a self-attention (SA) mechanism and a feed-forward network (FFN).
    \begin{itemize}
        \item \textbf{Self-Attention (SA):} For queries \( Q \), keys \( K \), and values \( V \), which are projections of \( X \), the attention function is computed as:
        \begin{equation}
            \text{Attention}(Q, K, V) = \text{Softmax}\left(\frac{QK^\top}{\sqrt{d_k}}\right)V
        \end{equation}
        where \( d_k \) is the dimension of the key.
        
        \item \textbf{Feed-Forward Network (FFN):} The FFN is applied to the output of the attention layer, defined as:
        \begin{equation}
            \text{FFN}(x) = \max(0, xW_1 + b_1)W_2 + b_2
        \end{equation}
    \end{itemize}
    Each layer includes skip connections and is followed by layer normalization.
    
    \item \textbf{Classifier Head:} The representation \( X_L[0] \) from the first token of the last layer \( L \) is passed through a linear layer to obtain the final predictions \( y \):
    \begin{equation}
        y = \text{Linear}(X_L[0])
    \end{equation}
\end{enumerate}

\subsection{Trainable Quantum Embedding with Classical Linear Transformation}
The trainable quantum embedding \cite{thumwanit2021trainable} with Transformer algorithm is an advanced approach integrating quantum computing principles with modern machine learning techniques. This hybrid algorithm at algorithm \ref{trainable_aglo} begins by taking images and their corresponding labels as inputs. It utilizes a Transformer model, a type of deep learning model renowned for its effectiveness in handling sequential data to embed each image into a high-dimensional space, effectively capturing the complex patterns and features within the pictures.

Once the images are embedded, they are reduced to a qubit-sized representation, aligning them with the requirements of quantum computing frameworks. This qubit representation is then processed through a Quantum Neural Network (QNN), which operates using the principles of quantum mechanics to perform computations that would be infeasible for classical neural networks. The QNN applies a quantum feature map \cite{schuld2019quantum, goto2020universal} and a quantum ansatz—parametric operations that encode the data into a quantum state and then transform it in a way that is dependent on the parameters being learned.

The output from the QNN is then measured, collapsing the quantum state into a classical form that can be used to compute a loss function. This loss function evaluates the accuracy of the model's predictions against the true labels, providing a basis for updating the Transformer and QNN parameters through backpropagation \cite{verdon2018universal, gonccalves2016quantum}. This algorithm adjusts parameters to minimize errors.

\subsubsection{Mathematical Model}
Linear layer transformation with trainable weights and the quantum embedding featuring two sets of Hadamard (\(H\)) and \(U_1(2\theta)\) gates (Z feature map), we will create a hybrid quantum-classical model using quantum kernel method. This model will utilize the output of the classical linear layer as the input parameter (\(\theta\)) for the quantum gates in the quantum embedding process.

The input to the model is a 1000-dimensional vector from the ViT trained for ImageNet-1K benchmark \( \mathbf{x} \). The linear layer transformation is given by:

\[ y = \mathbf{w}^T \mathbf{x} + b \]

\begin{itemize}
    \item \( \mathbf{w} \) is the trainable weight vector of the linear layer with 1000 elements.
    \item \( b \) is the trainable bias term.
    \item \( y \) is the scalar output of the linear layer, which will be used as the parameter \( \theta \) in the quantum embedding.
\end{itemize}

The scalar output \( y \) from the linear layer is used to parameterize the quantum gates in the embedding. The quantum state transformation is shown as the following.

\subsection{Quantum Feature Map and Ansatz}
Once the ViT processes the visual data, the linear representation layer transforms it. It embeds it into a quantum neural network shown in Figure \ref{fig:qnnarchitecture} to train the quantum neural network with the quantum Ansatz. The Pauli Feature Map, designed explicitly as a first-order Pauli Z-evolution circuit, stands as a cornerstone in quantum computing for embedding classical data into a quantum state. The Pauli Feature Map facilitates the construction of highly complex, high-dimensional feature spaces that classical computing methods struggle to replicate, opening new avenues for advancements in quantum machine learning and data analysis.

In the described quantum neural network including the Z-feature map and Real Amplitude Ansatz shown in Figure \ref{fig:qnnarchitecture}, the application of Hadamard (H) and U1 gates with a parameterized rotation \(2.0 \times x[0]\) transforms an initial quantum state \(|\psi\rangle\) through a series of operations that intricately encode information into the quantum system. Starting with the Hadamard gate, it places the qubit into a superposition, a fundamental state for quantum computation that allows a qubit to combine '0' and '1' states. This is mathematically represented as:

\begin{figure}[h]
\centering
\includegraphics[width=0.5\textwidth]{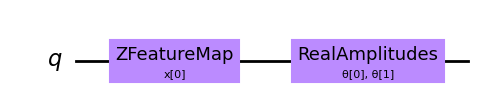}
\caption{Quantum neural network's architecture. Once the data processed by the transformer, the linear representation layer will transform the data and embed into the quantum circuit forming quantum state of information.}
\label{fig:qnnarchitecture}
\end{figure}

\begin{equation}
H = \frac{1}{\sqrt{2}}
    \begin{pmatrix}
    1 & 1 \\
    1 & -1
    \end{pmatrix}
\end{equation}

Following the superposition, the U1 gate introduces a phase shift dependent on the value \(2.0 \times x[0]\) without altering the probability amplitudes of the qubit states. The U1 gate is defined as:

\begin{equation}
U1(\lambda) =
    \begin{pmatrix}
    1 & 0 \\
    0 & e^{i\lambda}
    \end{pmatrix}
\end{equation}

where \( \lambda = 2 \times x[0] \).

The sequence of applying H, U1, H, and then U1 again results in a complex transformation of the initial state \(|\psi\rangle\), which is captured by the final formula:
\begin{equation}
 |\psi'\rangle = U1(2 \times x[0]) \cdot H \cdot U1(2 \times x[0]) \cdot H \cdot |\psi\rangle 
\end{equation}

This formula encapsulates the evolution of the qubit's state through the circuit, embedding the parameter \(x[0]\) into its phase and superposition, culminating in a state \(|\psi'\rangle\) that carries this encoded information. This process exemplifies the power of quantum circuits to manipulate quantum states, paving the way for sophisticated quantum algorithms that exploit these unique quantum phenomena for computational advantage.

The final quantum state \( |\psi(y)\rangle \) is obtained, where \( y \) is derived from the classical linear layer's output and measured in the computational basis, yielding probabilities \( P(0) \) and \( P(1) \) for the qubit being in the state \( |0\rangle \) and \( |1\rangle \), respectively. These probabilities are used to define the binary cross-entropy loss as the objective function:

\[ \mathcal{L} = -\left[y_{\text{true}} \log P(0) + (1 - y_{\text{true}}) \log P(1)\right] \]

where \( y_{\text{true}} \) is the binary label associated with the input \( \mathbf{x} \).

During training, the goal is to adjust the parameters of the classical linear layer (\( \mathbf{w} \) and \( b \)) to minimize the loss function \( \mathcal{L} \). This process involves backpropagation through the quantum circuit, which can be challenging due to the non-classical nature of quantum state transformations. Techniques such as the parameter shift rule may be employed to compute gradients of quantum circuits to classical parameters.

\begin{algorithm}
\caption{Trainable Quantum Embedding with Vision Transformer}
\label{trainable_aglo}
\begin{algorithmic}
\State \textbf{Inputs:}
\Statex $\mathcal{I}$: Set of images
\Statex $\mathcal{L}$: Corresponding labels
\State \textbf{Output:} Optimized model parameters

\Procedure{Trainable Quantum Embedding}{}
    \State Initialize parameters for Transformer and QNN
    \While{not converged}
        \For{each $(i, l) \in (\mathcal{I}, \mathcal{L})$}
            \State $e_i \gets$ Embed image $i$ using Transformer
            \State $q_i \gets$ Reduce $e_i$ to qubit size
            \State $d_i \gets$ Apply QNN to $q_i$ and measure
            \State $\mathcal{F} \gets$ Compute loss for $d_i$ and $l$
            \State Update parameters to minimize $\mathcal{F}$
        \EndFor
        \State Assess model on the validation set
    \EndWhile
\EndProcedure
\end{algorithmic}
\end{algorithm}

\section{Results}
The results from the empirical experimentation demonstrate that transformer-based quantum embedding is a stable and highly effective method for quantum neural networks using kernel methods. The transformer-based method outperforms the CNN-based variant by around 3 percent in the F1 score metric. Moreover, transformer-based quantum embedding delivers exceptional reliability and consistently high precision and accuracy with a standard deviation of 90 percent more.

\subsection{Effectiveness on Binary Classification}
\begin{figure}[h]
\centering
\includegraphics[width=0.5\textwidth]{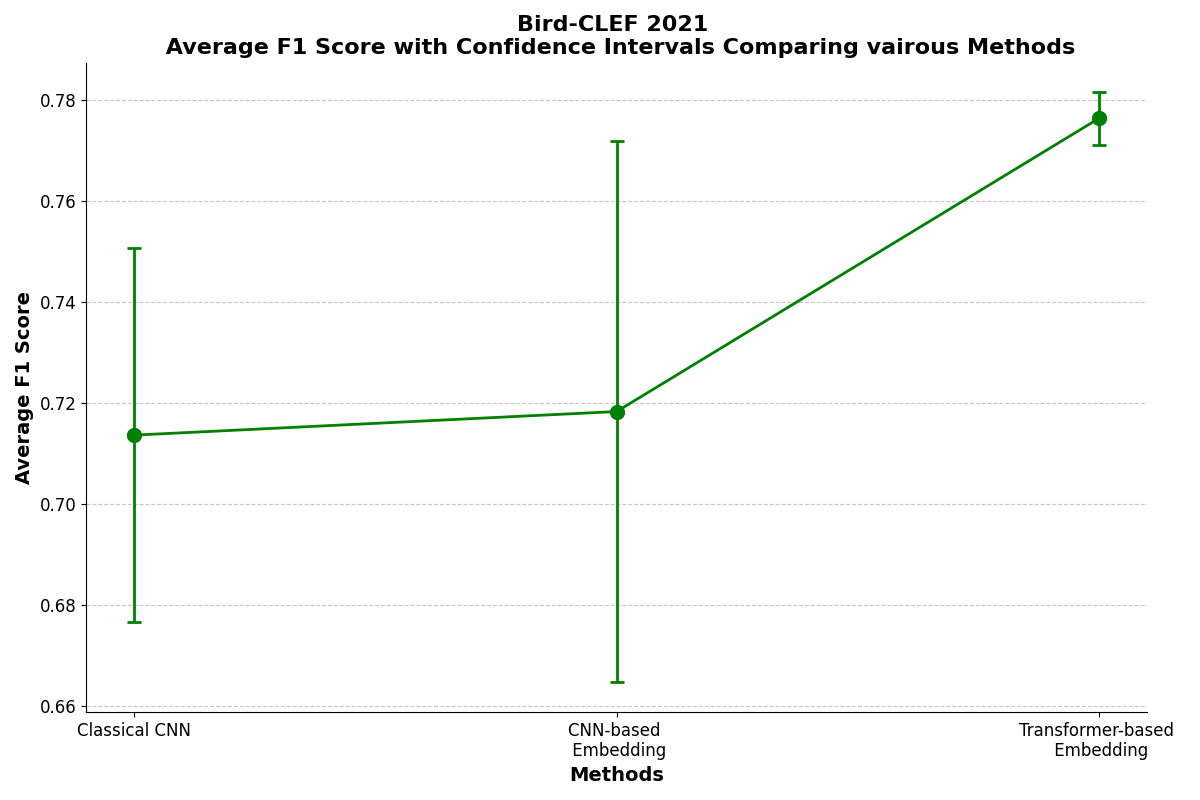}
\caption{Comparative results for the BirdCLEF-2021 dataset over various embedding methods, including classical CNN method (at the far left), CNN-based quantum embedding (at the middle), the architecture proposed in this research, which is transformer-based quantum embedding (at the far right)}
\label{fig:model_comparasion_birdclef}
\end{figure}

Figure \ref{fig:model_comparasion_birdclef} presents the results from the Bird-CLEF 2021 challenge, focusing on the comparative analysis of F1 scores, which measure the harmonic balance between precision and recall in binary classification tasks. The figure demonstrates that the transformer-based quantum embedding method exhibits remarkable effectiveness and reliability over the classical CNN and CNN-based embedding methods. The F1 scores for the Transformer-based architecture consistently outperform the other variants or alternatives, culminating in an average score of approximately 0.785, substantially higher than its counterparts.

\begin{table}[ht]
\centering
\caption{Comparison of Statistical Results from Bird-CLEF 2021 Binary Classification Over Various Methods}
\label{tab:comparison}
\begin{tabular}{@{}lcc@{}}
\toprule
 & \textbf{Standard Deviation} & \textbf{Median F1} \\ 
\midrule
Classical Baseline   & 0.0370 & 0.728 \\
CNN-based            & 0.0536 & 0.741 \\
Transformer-based    & \textbf{0.0052} & \textbf{0.774} \\
\bottomrule
\end{tabular}
\end{table}

\subsection{Performance Reliability}
Furthermore, the narrow confidence intervals at Figure \ref{fig:model_comparasion_birdclef} associated with the Transformer-based Quantum Embedding method indicate its reliability. Despite the increasing complexity of the classification tasks, the technique shows less variance in performance, suggesting that it is practical and robust to changes and challenges inherent in the Bird-CLEF dataset.

\begin{figure}[h]
\centering
\includegraphics[width=0.5\textwidth]{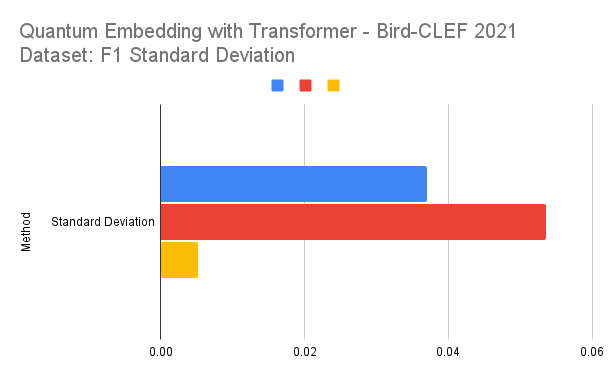}
\caption{Comparative results for the standard deviation of F1 score for BirdCLEF-2021 dataset over various embedding methods or training method, including classical CNN method (at the top), CNN-based quantum embedding (at the middle), the architecture proposed in this research, which is transformer-based quantum embedding (at the button)}
\label{fig:model_f1_sd_comparasion_birdclef}
\end{figure}

Figure \ref{fig:model_f1_sd_comparasion_birdclef} shows that our quantum embedding architecture with the ViT model outperforms all alternative methods by a significant level in terms of the standard deviation (SD) of the F1 score in the Bird-CLEF 2021 dataset. A significantly low SD demonstrates that the architectural innovation with a transformer could consistently yield stable and high-performing results with high precision and accuracy under binary classification.

In summary, the Transformer-based Quantum Embedding method stands out as both practical and reliable, outshining conventional and CNN-based methods in the high-stakes arena of bio-acoustic event detection, as reflected by its superior and consistent F1 scores in the Bird-CLEF 2021 challenge. This evidence implies that the architectural innovation of this study has yielded a significant advancement in shallow quantum circuits to deliver promising machine learning capability with an effective and efficient embedding scheme.

\section{Discussion}
\subsection{Extension to Larger Circuits}
The architecture proposed in this study is extensible to larger quantum circuits with more than one qubit. The linear transformation layer is flexible and adaptable to various circuit sizes. However, the effectiveness of embedding requires further investigation to ensure the overall effectiveness of this embedding scheme.

\subsection{Technical Challenges} 
The trainable method in this study involves gradient descent and backpropagation combined with classical and quantum elements, which might encounter risks like a barren plateau or hybrid gradient descent. Further investigation into such challenges and understanding of this architecture's theoretical and practical constraints are critical.

\subsection{Outlooks}
Considering the substantial advancement of this architectural innovation in quantum embedding, the architecture paves the way for future quantum kernel methods with quantum neural networks (QNNs) to model sequential or recurrent data patterns with the aid of transformer-based models. This could be a new pathway for quantum machine learning in solving complex computer vision (CV) and natural language processing (NLP) tasks.

\section{Conclusion}
The research proposed an architectural innovation to facilitate quantum embedding with a transformer model for solving high-dimensional binary classification problems using the single-qubit classifier. The empirical evidence has shown that our architectural innovation from quantum embedding to classifier design ha s significantly improved compared to the classical counterpart and another quantum embedding variant. The research implies that our embedding architecture with a transformer is a better, more reliable scheme for facilitating quantum machine learning in solving complex problems.




\nocite{langley00}

\bibliography{example_paper}
\bibliographystyle{icml2024}

\end{document}